\documentclass{an}
\usepackage{graphicx}
\usepackage{times}
\usepackage{fancyhdr}
\begin{document}

\title{Ultra low-mass star and substellar formation in $\sigma$ Orionis}

\author{J. A. Caballero\thanks{On behalf of the JOVIAN Collaboration.
The members are R. Rebolo, V.J.S B\'ejar, E.L. Mart\'{\i}n, J.A. Caballero
(Instituto de Astrof\'{\i}sica de Canarias), M.R. Zapatero Osorio, D. Ba\-rra\-do y
Navascu\'es (Laboratorio de Astrof\'{\i}sica Espacial y F\'{\i}sica
Fundamental--INTA), R. Mundt, C.A.L. Bailer-Jones (Max-Planck-Institut f\"ur
Astronomie), J. Eisl\"offel (Th\"uringer Ladessternwarte Tautenburg) and T.
Forveille (Canada--France--Hawai'i Telescope).}}  
\institute{Instituto de Astrof\'{\i}sica de Canarias}

\date{Received; accepted; published online}

\abstract{
The nearby young $\sigma$ Orionis cluster ($\sim$360 pc, $\sim$3 Ma) is becoming
one of the most important regions for the study of ultra low-mass star formation and
its extension down to the mass regimes of the brown dwarfs and 
planetary-mass objects. 
Here, I introduce the $\sigma$ Orionis cluster and present three studies that the
{JOVIAN} group is developing: 
a pilot programme of near-infrared adaptive-optics imaging,
intermediate-resolution optical spectroscopy of a large sample of stars of the
cluster and 
a study of the mass function down to the planetary-mass domain.
This paper is a summary of the content of four posters that I presented in the
{\em Ultra low-mass star formation and evolution} Workshop, as single author or
on behalf of different collaborations.
} 

\correspondence{zvezda@iac.es}

\maketitle

\section{The $\sigma$ Orionis cluster}

Known for centuries, the star $\sigma$ Orionis in the Ori OB1b Association (the
Orion Belt), close to the Horsehead Nebula, gives the name to the open cluster
that surrounds the star.  
It is, in reality, a quintuple system (at least) of OB-type stars that
injects energy and turbulence into the interstellar medium.
Garrison (1967) and Lyng\aa~(1981) were the first authors to recognize the
$\sigma$ Orionis cluster. 
The existence of a population of low-mass stars has been known since the end of the
nineties, when Wolk (1996) and Walter, Wolk \& Sherry (1998) discovered an
overdensity of X-ray sources in the region. 
Many studies have been performed there, from the search for strong H$\alpha$
emitters (Haro \& Moreno 1953; Wiramihardja et al. 1989) to the 
detection of objects below the deuterium burning mass limit (Zapatero Osorio et
al. 2000), through the determination of the mass function in the substellar
domain (B\'ejar et al. 2001), 
photometric variability of brown dwarfs and very-low-mass stars (Scholz \&
Eisl\"offel 2004), 
or the characterization in the mid-infrared of discs of Classical T Tauri stars
(Oliveira, Jeffries \& van Loon 2004). 
The cluster has the most suitable properties for searching and characterizing its
rich substellar population: it is young (3$^{+5}_{-1}$ Myr; Zapatero Osorio et
al. 2002), nearby (360$^{+70}_{-60}$ pc; Brown et al. 1994) and has a low
reddening ($A_V <$ 1; Lee 1968).  
A complete review of the $\sigma$ Orionis cluster is found in Caballero (2005).

\begin{figure}
\resizebox{\hsize}{!}
{\includegraphics[]{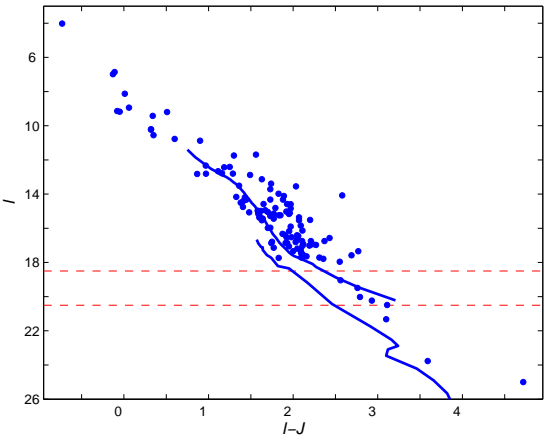}}
\caption{An $I$ vs. $I-J$ colour--magnitude diagram showing most of the cluster
members with spectroscopic features of youth (Li {\sc i} in absorption, strong
and asymmetric H$\alpha$ emission, weak alkali lines, etc.).
It covers 21 mag in $I$-band magnitude or four orders of magnitude in mass, with
objects from 20 solar masses to 3 Jupiter masses.
Member candidates have been gathered from the literature (e.g. Zapatero Osorio
et al. 2002; Kenyon et al. 2005).
Dashed horizontal lines denote the hydrogen (top) and deuterium (bottom) burning
limits.
NextGen and Cond isochrones for 3 Myr, roughly valid in the low-mass stellar and
substellar domains, respectively, are also shown in solid lines (Baraffe et al.
1998, 2003).
All new objects presented in this paper fall in the cluster sequence.
} 
\label{fig_0}
\end{figure}

Here I present preliminary results on three studies performed within the JOVIAN
Collaboration aimed at shedding light on the substellar formation mechanisms.
They are focused on the relationship between the stellar and substellar
populations (Sections \ref{sO_III} and \ref{sO_I}) and the investigation of the
latter down to the planetary-mass domain (Section \ref{sO_II}).
Further details will be given in forthcoming papers.

\section{Adaptive-optics imaging of stellar cluster members}
\label{sO_III}

Substellar objects, when companions to stars, are found at distances between
$\sim$50 and $\sim$3\,600 AU of the primaries (e.g. Nakajima et al. 1995, Rebolo
et al. 1998).
While many multiple stellar systems and isolated substellar objects are found in
the $\sigma$ Orionis cluster, no brown dwarf or planetary-mass object has been
detected yet at projected physical separations from stellar members at less than
about 10\,000 AU.
Through a pilot programme of near-infrared adaptive-optics (AO) imaging of six
stellar cluster members, we have investigated the corona between $\sim$150 and
$\sim$7\,000 AU from the primaries.

\subsection{Observations and reduction}

During a campaign in 2003 Sep, we took deep high spatial-resolution
$H$-band images of six stellar members of the $\sigma$ Orionis cluster.
We used the {NAOMI+INGRID} AO system at the 4.2 m William Herschel
Telescope, which offers a field of view of 41 $\times$ 41 arcsec$^2$ with a pixel size of
about 0.0386 arcsec\,pixel$^{-1}$.
The FWHMs of the images after AO correction was between 0.4 and 0.2 arcsec.
Near-infrared images were sky-subtracted, flat-fielded, aligned and combined
within the {\sc iraf} environment.
We reached magnitudes that would allow us to detect planetary-mass
cluster candidates.

\begin{figure}
\resizebox{\hsize}{!}
{\includegraphics[]{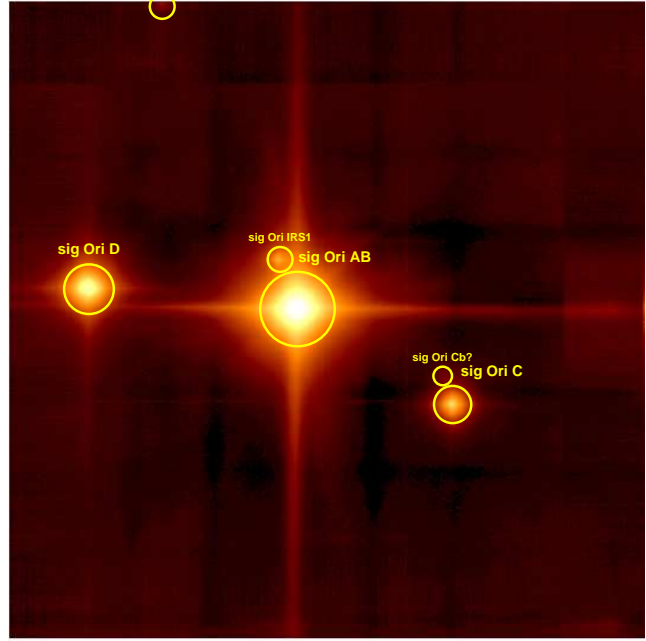}}
\caption{NAOMI+INGRID $J+H$-band image of $\sigma$ Ori AB and visual
companions.
Several objects are marked with a circle.
North is up, east is to the left. 
Field of view is 41 $\times$ 41 arcsec$^2$.
$\sigma$ Orionis IRS1 is 3.3 arcsec to the north-northeast of $\sigma$ Orionis
AB.
}
\label{fig_III}
\end{figure}

\begin{table}[h]
\caption{}
\label{table_sO_III}
\begin{tabular}{lcr}
\hline
Name 		& Sp. type	& Notes	\\
\hline
$\sigma$ Ori AB & O9.5V$+$B0.5V & Multiple, 1 new comp.	\\
HD 294271 	& B5V		& 2 new comp.		\\
HD 37525 	& B5V		& Double, 1 PMS comp.	\\
HD 37686 	& B9Vn		& Dust shell, 1 new comp.\\
HD 37333 	& A1Va		& \\
4771--899 	& K7.0		& Triple		\\
\hline
\end{tabular}
\end{table}

\subsection{Results}

The observed stars cover a wide range of spectral types, from O9.5V to K7.0. 
Their names and spectral types are shown in Table \ref{table_sO_III}. 
Apart from the AO images, we have used other near infrared, optical ($VRI$) and
X-ray data to derive the real astrophysical nature of the detected visual
companions.
A total of 22 visual companions to the primary targets have been detected in
this pencil-beam survey.
Six sources show blue optical-near infrared colours for their magnitudes, and
they do not match in any colour--magnitude diagram of the cluster.
There is not enough information to derive the nature of other five sources
(including a faint object $\sim$2 arcsec northeast of $\sigma$ Ori C,
labelled in Fig. \ref{fig_III}).
Eleven objects remain as cluster member candidates according to their
magnitudes and colours.

\begin{itemize}

\item Three of them were previously known cluster members: $\sigma$ Ori C,
$\sigma$ Ori D (surrounding $\sigma$ Ori AB) and S\,Ori J053847.5--022711 (close
to [W96] 4771--899).

\item We have detected the near-infrared counterpart of the mid-infrared and
radio source $\sigma$ Ori IRS1, a dust cloud next to $\sigma$ Ori AB discovered
by van Loon \& Oliveira (2003). 
We confirm the claim by Sanz Forcada et al. (2004) that it is also an X-ray
emitter.
The object is detected in {\em Chandra} archive images taken with the HRC-I
instrument (see Fig. \ref{fig_III}). 

\item One of the HD 37525 companions seem to be the pre-main sequence
photometric candidate star P053902--0238, discovered by Wolk (1996).

\item Two bright objects are the previously unknown secondaries of the HD 37525
and [W96] 4771--899 close binary systems, at angular separations of
0.45$\pm$0.04 and 0.40$\pm$0.08 arcsec, respectively.  

\item The four remaining objects are visual companions to $\sigma$ Ori AB (1),
HD 294271 (2) and HD 37686 (1) at separations in the range 5.5--19.0 arcsec.
The object close to $\sigma$ Ori AB, also an X-ray emitter, and the outer HD
294271 companions have been detected in previous surveys (Caballero et al., in
prep. and Section \ref{sO_I} of this paper, respectively).
They and the inner HD 294271 companion are probably early-to-intermediate M-type
very low-mass stars of the cluster. 
The extremely red object ($J-H \sim$ 1.9 mag) at 5.5 arcsec to HD 37686, a
B9Vn-type star with a dust shell and only a few arcmin away from the Horsehead
Nebula, could be either a very low mass star or a brown dwarf in the preliminary
stages of its evolution. 

\end{itemize}

\section{A bridge between the stellar and the substellar populations}
\label{sO_I}

The $\sigma$ Orionis cluster has become a unique laboratory for understanding
the processes that originate brown dwarfs and isolated planetary-mass objects.
In order to understand the formation process of substellar objects and the
relationship between these and stars, it is necessary to determine first the
stellar population of the cluster and its properties 
(spatial distribution, binary and disc frequency, etc.).

\subsection{Photometric survey, target selection and wide-field multifibre
spectroscopy} 

We performed shallow optical observations with the Wide Field Camera (0.333
arcsec\,pixel$^{-1}$) at the 2.5 m Isaac Newton Telescope in 2003 Jan.
We covered in $VRI$ bands more than 1 deg$^2$ in four different pointings around
the centre of the $\sigma$ Orionis cluster.
We subtracted the bias and divided by their respective flat-field images, and
performed aperture and PSF photometry using common tasks within the {\sc iraf}
environment.
We pre-selected several hundred sources from different colour--magnitude diagrams
with photometric data in the $VRI$ bands and magnitudes in the range 11 $\le I
\le$ 18.
$JHK_{\rm s}$-band follow-up photometry, using Two Mass All-Sky Survey data,
allowed us to select the list of cluster star and massive brown-dwarf candidates
to be observed spectroscopically.

\begin{figure}
\resizebox{\hsize}{!}
{\includegraphics[]{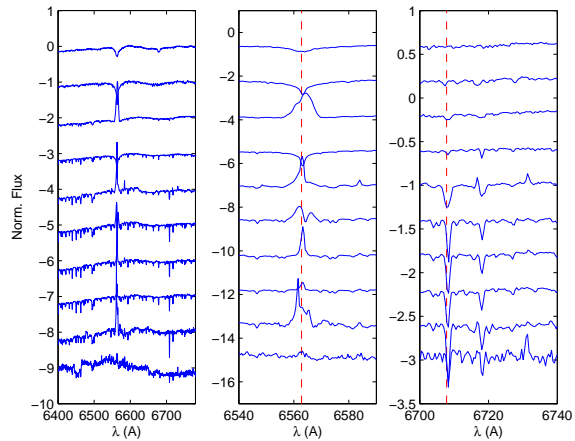}}
\caption{Example of ten spectra of cluster members taken with {WYFFOS+AF2}.
The whole $\lambda$ coverage is shown to the left, while the H$\alpha$/[N {\sc
ii}] and the Li {\sc i}/[S {\sc ii}] regions are shown in the centre and to the
right, respectively.
Spectral type increases from top (B2V, $\sigma$ Orionis D) to bottom (intermediate
M).}
\label{fig_I}
\end{figure}

In 2003 Nov, we used the Wide Field Fibre Optical Spectrograph instrument and
the robot positioner AutoFib2 ({WYFFOS+AF2}) at the 4.2 m William Herschel
Telescope to obtain about 200 intermediate-resolution (R $\sim$ 8\,000; nominal
resolution of 0.43 \AA\,pixel$^{-1}$) spectra of sources in the direction of
$\sigma$ Orionis. 
We covered the wavelength range between 6\,400 and 6\,800 \AA. 
The {\sc dohydra} task was used during the extraction, sky subtraction and
wavelength calibration of the spectra.

\subsection{Results}

We compiled a list of 80 members of the $\sigma$ Orionis cluster
with {WYFFOS+AF2} spectroscopy, based on the presence of Li {\sc i}
$\lambda$6707.8 \AA~in absorption and H$\alpha$ in emission (mid- and late-type
stars) or spectral type determination (early type).
About one half out of the objects are spectroscopically studied here for the
first time. 
Using available data on the members, we have investigated:

\begin{itemize}

\item the variation of the strength of the Li {\sc i} with spectral type and
with signal-to-noise ratio (from 0.05 \AA~in late F stars to 0.70 \AA~in
intermediate M stars);

\item the frequency of accretors according to the White \& Basri (2003)
criterion (46$^{+16}_{-13}$\% of K and M stars) and the presence of asymmetries
in the profiles of the H$\alpha$ line;

\item the existence of forbidden lines in emission ([N {\sc ii}]
$\lambda\lambda$6548.0,6583.5, [S {\sc ii}] $\lambda\lambda$6716.4,6730.8);

\item the widening of photospheric lines (of up to 100 km\,s$^{-1}$);

\item the relationship between the $L'$- and $K_{\rm s}$-band flux excesses and
the spectroscopic features associated with accretion from protoplanetary discs;

\item the average of the radial velocity of the cluster members ($+$30.2
km\,s$^{-1}$), the standard deviation with respect to the mean radial velocity
corrected of systematic dispersion ($\sim$2.4 km\,s$^{-1}$) and the existence of
radial velocity outliers (probably due to unresolved close companions);

\item and the frequency of X-ray emitters catalogued by {\em ROSAT} and {\em ASCA}
space observatories as a function of spectral type (the bulk of the K stars are
X-ray emitters);


\end{itemize}

The sample has also allowed us to create a complete database of stellar members
of the $\sigma$ Orionis cluster, necessary for studying the spatial distribution of
the member stars and to find the typical distance between these and
the substellar objects detected in deep surveys.

\section{Mass function down to the planetary-mass domain}
\label{sO_II}

The main goal of the JOVIAN Collaboration is to characterize the mass
function in the substellar domain and to search for objects that do not burn
deuterium and have masses of less than 13 Jupiter masses ($M_{\rm Jup}$), the
boundary between brown dwarfs and planetary-mass objects.
Our team has performed a new very deep photometric survey in the $IJ$ bands to
the southeast of the centre of the $\sigma$ Orionis cluster, covering the
whole brown-dwarf and part of the planetary-mass domains.

\subsection{The $IJ$-band survey and the $HK$-band follow up}

We have obtained photometric data using the Wide Field Camera at the 2.5 m Isaac
Newton telescope in the $I$ band (2000 Dec) and with the {ISAAC}
spectrograph and camera (0.148 arcsec\,pixel$^{-1}$) at the Very Large Telescope
UT1 in $J$ band (2001 Dec).
The overlapping area between the optical and the near infrared images was 790
arcmin$^2$ (there is an overlap of $\sim$80\% between this search and that
presented in Sect. \ref{sO_I}).
Completeness and limiting magnitudes in each set of images were 23.4 and 24.0,
and 20.6 and 21.8 mag, respectively.
We have reduced the images, perfomed aperture and PSF photometry (using standard
{\sc iraf} tasks) and searched the optical counterparts of the near-infrared
sources.
After the magnitude and astrometric calibrations, $I$- and $J$-band magnitudes
and equatorial coordinates for about 10\,000 sources were available.
We selected new cluster member candidates in the $I$ vs. $I-J$ diagram in the
same way as in B\'ejar et al. (2004).

In order to verify that the $IJ$-selected objects had near infrared colours
typical of late M and L dwarfs, and to discriminate them from reddened stellar
sources or high-redshift galaxies, we used the instruments {OMEGA-2000} at
the 3.5 m Calar Alto Teleskop and {CHFT-IR} at the 3.6 m
Canada-France-Hawai'i Telecope to perform an $HK$-band follow up.

\subsection{Results}

A total of 39 cluster member candidates in the substellar mass domain have
arised from our deep survey and follow up.
They have $I$- and $J$-band magnitudes in the ranges 16.5 to 22.8 and 14.5 to
19.5, respectively.
Out of them, 30 objects had been detected in previously published surveys.
We have spectroscopic data for 25 objects, most of them showing spectroscopic
features common to very young low-mass objects.

We have derived a mass spectrum ($dN/dM \propto M^{-\alpha}$) after computing
the mass for each substellar object using the mass-luminosity relation offered
by theoretical models of the Lyon group and the relation between $BC_J$
bolometric correction and the $I-J$ colours of the field ultracool dwarfs with
known parallax.
The most probable masses for each object range between 74 and 6 $M_{\rm Jup}$.
Eight out of the 39 cluster member candidates have masses below the deuterium
burning mass limit.
The slope of the mass spectrum in this mass range is $\alpha \equiv -\gamma =
+0.3 \pm 0.2$.
The study of the stellar mass spectrum in this area will be presented in
Caballero (2005).

\begin{figure}
\resizebox{\hsize}{!}
{\includegraphics[]{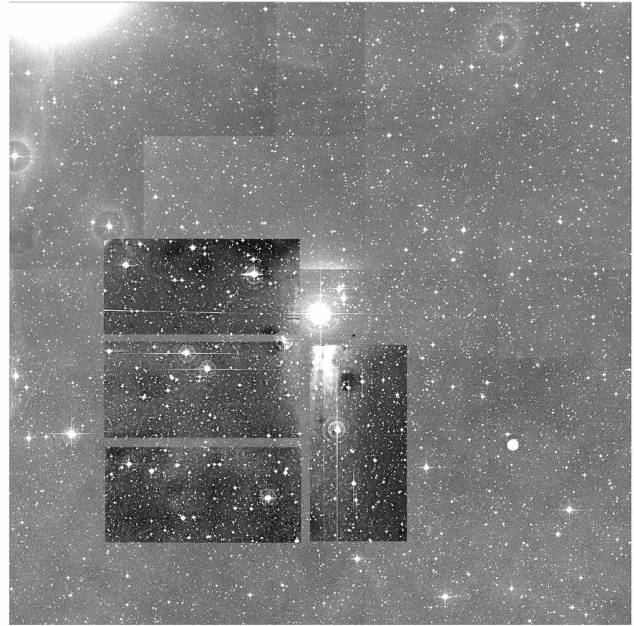}}
\caption{$I$-band WFC images on to Digitised Sky Survey--2--IR mosaic
centred in $\sigma$ Orionis AB.
North is up and east to the left.
Size of each WFC chip is about 11 $\times$ 22 arcmin$^2$.}
\label{fig_II}
\end{figure}

\section{Conclusions}

We have gone deeper in the search for objects with masses close to the opacity
limit for fragmentation in the nearby young $\sigma$ Orionis cluster and have
performed different photometric, both optical and infrared, and spectroscopic
studies with the aim of characterizing the cluster stellar and substellar
populations and how they are connected.






\begin{thebibliography}{}
\bibitem{Ba98} Baraffe, I., Chabrier, G., Allard, F., Hauschildt, P.H.: 1998,
A\&A 337, 403
\bibitem{Ba03} Baraffe, I., Chabrier, G., Barman T.S., Allard, F., Hauschildt,
P.H.: 2003, A\&A 402, 701
\bibitem{Be01} B\'ejar, V.J.S., Mart\'{\i}n, E.L., Zapatero Osorio, M.R.,
Rebolo, R., Ba\-rra\-do y Navascu\'es, D. et al.:  
2001, ApJ 556, 830
\bibitem{Be04} B\'ejar, V.J.S., Caballero, J.A., Rebolo, R., Zapatero Osorio,
M.R., Ba\-rra\-do y Navascu\'es, D.: 2004, ApSS 292, 339  
\bibitem{Br94} Brown, A.G.A., de Geus, E.J., de Zeeuw, P.T.: 1994, A\&A 289, 101
\bibitem{C05} Caballero, J.A.: 2005, Ph.D. thesis, Universidad de La Laguna
\bibitem{G67} Garrison, R.F.: 1967, PASP 79, 433
\bibitem{HM53} Haro, G., Moreno, A.: 1953, BOTT 1g, 11
\bibitem{K05} Kenyon, M.J., Jeffries, R.D., Naylor, T., Oliveira, J.M., Maxted,
P.F.L.: 2005, MNRAS 356, 89
\bibitem{Le68} Lee, T.A.: 1968, ApJ 152, 913
\bibitem{Ly81} Lyng\aa, G.: 1981, ADCBu 1, 90
\bibitem{Na95} Nakajima, T., Oppenheimer, B.R., Kulkarni, S.R., Golimowski,
D.A., Mathews, K., Durrance, S.T.: 1995, Nature 378, 463
\bibitem{OJvL04} Oliveira, J.M., Jeffries, R.D., van Loon, J.Th.: 2004, MNRAS 347,
1327
\bibitem{Re95} Rebolo, R, Zapatero Osorio, M.R., Madruga S., B\'ejar, V.J.S.,
A\-rri\-bas, S., Licandro, J.: 1998, Nature 377, 129
\bibitem{SE04} Scholz, A., Eisl\"offel, J.: 2004, A\&A 419, 249
\bibitem{SFFP04} Sanz-Forcada, J., Franciosini, E., Pallavicini, R.: 2004, A\&A 
421, 715
\bibitem{vLO} van Loon, J.Th., Oliveira, J.M.: 2003, A\&A 405L, 33
\bibitem{W96} Wolk, S.J.: 1996, Ph.D. thesis, State Univ. New York at Stony Brook
\bibitem{WWS98} Walter, F.M., Wolk, S.J., Sherry, W.: 1998, ASP Conf. Ser. 154, 1793
\bibitem{WB03} White, R.J.; Basri, G.: 2003, ApJ 582, 1109
\bibitem{W89} Wiramihardja, S.D., Kogure, T., Yoshida, S., Ogura, K., Nakano,
M.: 1989, PASJ 41, 155
\bibitem{ZO00} Zapatero Osorio, M.R., B\'ejar, V.J.S., Mart\'{\i}n, E.L.,
Rebolo, R., Ba\-rra\-do y Navascu\'es, D. et al.:  
2000, Sci 290, 103 
\bibitem{ZO02} Zapatero Osorio, M.R., B\'ejar, V.J.S., Pavlenko, Ya., Rebolo,
R., A\-lle\-nde Prieto, C. et al.: 
2002, A\&A 384, 937
\end{thebibliography}
\end{document}